\documentclass[preprint]{revtex4}
\usepackage{amsmath}
\usepackage{graphicx}
\usepackage{amssymb}
\usepackage{amsthm, amscd}
\usepackage{amsfonts}
\usepackage{cancel}
\usepackage{times}
\usepackage{pstricks}
\usepackage{wasysym}
\usepackage{ulem}
\usepackage[dvips]{epsfig}
\usepackage{subfigure}
\usepackage{amssymb}

\newcommand{\dslash}{\not{\hbox{\kern-2pt $\partial$}}}
\newcommand{\td}{\tilde}
\newcommand{\bq}{\begin{equation}}
\newcommand{\eq}{\end{equation}}
\newcommand{\bqa}{\begin{eqnarray} \displaystyle}
\newcommand{\eqa}{\end{eqnarray}}
\newcommand{\nn}{\nonumber \\}
\newcommand{\bw}{\begin{widetext}}
\newcommand{\ew}{\end{widetext}}

\newcommand{\JJ}{{\cal J}}
\newcommand{\tr}{\mbox{tr}}

\newcommand{\CCC}{{\{C_1,..,C_n\}}}
\newcommand{\CCCB}{{\{\bar C_1,..,\bar C_n\}}}
\newcommand{\CCCP}{{\{C_1^{'},..,C_n^{'}\}}}
\newcommand{\DCCC}{{\{T[C_1],..,T[C_n]\}}}
\newcommand{\DCCCP}{{\{T[C_1^{'}],..,T[C_n^{'}]\}}}
\newcommand{\CCCT}{{\{\td C_1,..,\td C_m\}}}
\newcommand{\DNCCC}{{\{ T^l[C_1],..,T^l[C_n] \}}}
\newcommand{\HK}{{(l)}}
\newcommand{\HKP}{{(l+1)}}

\begin{document}


\title{Holographic description of large $N$ gauge theory}

\author{Sung-Sik Lee$^{1,2}$\vspace{0.7cm}\\
{\normalsize{$^1$Department of Physics $\&$ Astronomy, McMaster University,}}\\
{\normalsize{1280 Main St. W., Hamilton ON L8S4M1, Canada}}\vspace{0.2cm}\\
{\normalsize{$^2$Perimeter Institute for Theoretical Physics,}}\\
{\normalsize{31 Caroline St. N., Waterloo ON N2L2Y5, Canada}}}

\date{\today}

\begin{abstract}

Based on the earlier work [S.-S. Lee, Nucl. Phys. B {\bf 832}, 567 (2010)],
we derive a holographic dual for 
the $D$-dimensional $U(N)$ lattice gauge theory
from a first principle construction.
The resulting theory is a 
lattice field theory of closed loops, 
dubbed as {\it lattice loop field theory}
which is defined on a $(D+1)$-dimensional space.
The lattice loop field theory
is well defined non-perturbatively,
and it becomes weakly coupled and local 
in the large $N$ limit with a large 't Hooft coupling.

\end{abstract}

\maketitle

\section{Introduction}

Although there are many non-trivial evidences for 
the anti-de Sitter space/conformal field theory (AdS/CFT) correspondence\cite{MALDACENA,GUBSER,WITTEN},
a first principle derivation of the conjecture is not available yet.
Nonetheless the correspondence has been employed to
a wide range of physical systems including 
condensed matter systems\cite{Hartnoll,Herzog,McGreevy,Sachdev},
in a hope that certain features of strongly
coupled quantum field theory can be 
captured by holographic theories constructed
based on phenomenological reasoning.
On the one hand, such approaches have produced
interesting insights into strongly coupled quantum many-body systems.
On the other hand, there exists a clear limitation
because it has not been possible to identify the precise holographic
dual for a general quantum field theory that one may want to understand.

Since the 't Hooft's observation that large $N$ gauge theory 
is related to a weakly interacting string theory\cite{tHOOFT},
it has been suggested that there exist 
intrinsic connections between the two theories\cite{POLYAKOV99}.
The fundamental object that bridges between these two theories
is Wilson loop which becomes 
classical in the large N limit.
The dual string theory that governs the classical 
equation of motion of Wilson loop\cite{POLYAKOV80,MM} in the loop space
is expected to be 
defined on one higher dimensional space
than the space on which the field theory is defined, 
and the additional dimension corresponds to the
energy scale in the renormalization group (RG) sense\cite{POLYAKOV99}.
There have been many works which 
made the connection between RG flow
of general quantum field theory and holographic theory
more precise\cite{EMIL,VERLINDE,LI,DAS,Gopakumar:2004qb,POLCHINSKI09,SLEE10,HEEMSKERK10,Faulkner1010}.

In this paper, we present a first principle construction 
of a holographic theory dual to the $U(N)$ gauge theory 
based on the prescription\cite{SLEE10}
which has been applied to the $O(N)$ vector model\cite{KLEBANOV,VASILIEV,PETKOU,GIOMBI,KOCH}.
We use the lattice regularization for the gauge theory. 
The derived holographic theory is
a lattice field theory of closed loop 
defined on a $(D+1)$-dimensional 
lattice which can be viewed as a 
discrete AdS space.
The holographic theory becomes classical in the large $N$ limit
and locality emerges when the 't Hooft coupling is large.
This construction provides a realization of an earlier idea of reformulating large N gauge theory as a classical theory
of gauge neutral fields\cite{WITTEN1980,COLEMAN}.

\section{From U(N) gauge theory to loop field theory}

We start with the $U(N)$ gauge theory
defined on the D-dimensional Euclidean hypercubic lattice,
\bqa
Z[\JJ] = \int d U e^{-S[U;\JJ]}
\eqa
with the action,
\bqa
S[U;\JJ] = - \sum_{n=1}^\infty  \sum_\CCC N^{2-n} \JJ_\CCC \prod_{i=1}^n W_{C_i}.
\eqa
Here $W_C$ is Wilson line defined on closed oriented loop $C$,
\bqa
W_C & = & \tr \left[ \prod_{<ij> \in C} U_{ij} \right],
\label{WC}
\eqa
where $U_{ij} = U_{ji}^\dagger$ is $U(N)$ matrix (holonomy)
defined on nearest neighbor bond $<i,j>$,
and $dU \equiv \prod_{<i,j>} dU_{ij}$.
In Eq. (\ref{WC}) and all products of holonomies hereafter,
we assume that the product is path-ordered
along the orientation of the curve.
$\sum_\CCC$ is the sum over unordered 
sets of $n$ closed loops.
$\JJ_{\{C_1\}}$, $\JJ_{\{C_1,C_2\}}$, ... are
coupling constants associated with
single-trace,
double-trace operators,
and so on.
For single trace coupling, 
we will also use the notation $\JJ_{C_1} \equiv \JJ_{\{C_1\}}$ interchangeably.
The factor of $N^{2-n}$ has been singled out from each coupling.
Roughly, the inverse of $\JJ_\CCC$ corresponds to
the 't Hooft coupling.
Throughout the paper, we will focus on
the large $N$ limit with fixed $\JJ_\CCC$.
In this limit, the action is manifestly proportional to $N^2$.
To guarantee the reality of the action,
we impose $\JJ_{\{\bar C_1, \bar C_2, ...\}} = \JJ_{\{C_1, C_2, ...\}}^*$
where $\bar C_i$ is the inverse loop of $C_i$, namely
the loop with the same trajectory but with the inverse orientation.
If one ignores all multi-trace deformations
and loops larger than the unit plaquette,
one restores the standard lattice gauge theory\cite{WILSON}.
Here we consider the most general gauge invariant action.
We assume that couplings associated with large loops
or multi-trace couplings with loops which are far from each other
are exponentially small in the size of large loops or
in the separation between loops.
If couplings satisfy this condition,
we say the theory is local.
We note that 
effective theories obtained by integrating
out short distance fluctuations generically
contain non-local terms
which are exponentially small.

\begin{figure}[h!]
\centering
        \includegraphics[height=4cm,width=8cm]{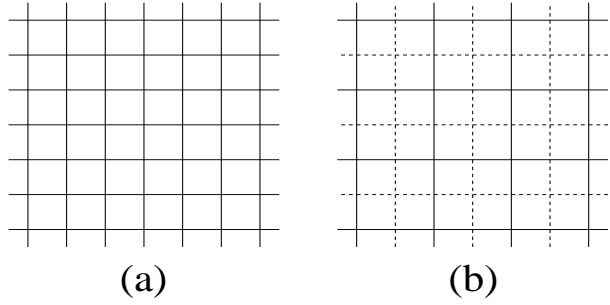}
\caption{(a) Two-dimensional illustration of the $D$-dimensional
hypercubic lattice.
(b) Links in the hypercubic lattice 
divided into two sets 
where solid links belong to the coarse grained lattice $X$
which form a hypercubic lattice with a larger lattice spacing
and the dashed links belong to $Y = X^c$.
}
\label{fig:lattice}
\end{figure}

\begin{figure}[h!]
\centering
        \includegraphics[height=4cm,width=5cm]{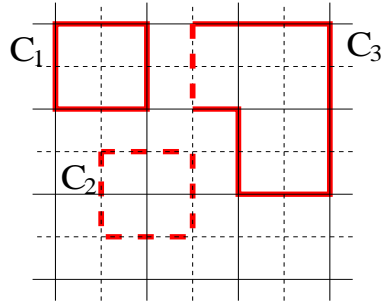}
\caption{
Three types of loops;
Wilson loops for $C_1$, $C_2$ and $C_3$ are included in 
$S_X$, $S_Y$ and $S_{XY}$, respectively.
}
\label{fig:loops}
\end{figure}

To set the stage for a real-space renormalization,
we divide links in the lattice into two sets\cite{FRADKIN79}.
The first set $X$ contains links that form a coarse grained lattice, that is,
the hypercubic lattice with the lattice spacing twice larger than
the original one.
The second set $Y$ contains all other links.
This is illustrated in Fig. \ref{fig:lattice}.
The action can be divided as
\bqa
S[U,\td U;\JJ] = S_X[U] + S_Y[\td U] + S_{XY}[U,\td U],
\eqa
where holonomy on links in $Y$ are denoted with tilde :
$U_{ij} \rightarrow \td U_{ij}$ if $<i,j> \in Y$.
Here $S_X$ and $S_Y$ are the actions 
which contain Wilson loops 
only in $X$ and $Y$, respectively,
and $S_{XY}$ includes loops that span across $X$ and $Y$,
as is shown in Fig. \ref{fig:loops}.
Now the partition function is written as
\bqa
Z[\JJ] 
& = & \int d U e^{-S_X[U]} \left< e^{- S_{XY}[U, \td U]} \right>_Y,
\eqa
where
$
\left< O \right>_Y = \int d \td U O e^{-S_Y[\td U]}
$.


\begin{figure}[h!]
\centering
        \includegraphics[height=4cm,width=14cm]{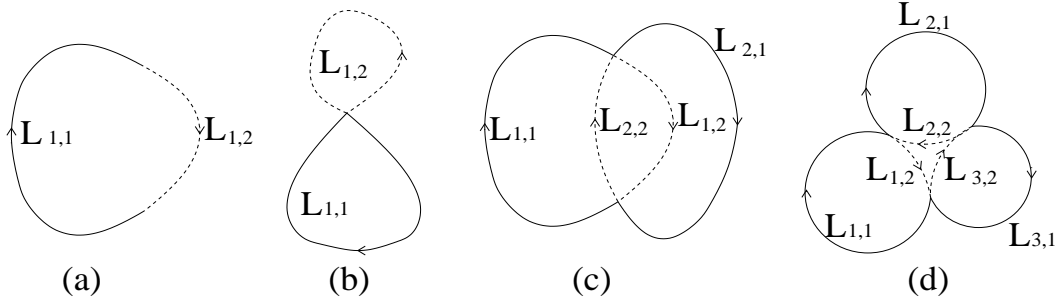}
\caption{
Examples of Wilson loops included in $e^{-S_{XY}}$.
The solid line represents links in $X$ 
and the dashed line in $Y$.
}
\label{fig:loops_examples}
\end{figure}

\begin{figure}[h!]
\centering
        \includegraphics[height=3.5cm,width=9cm]{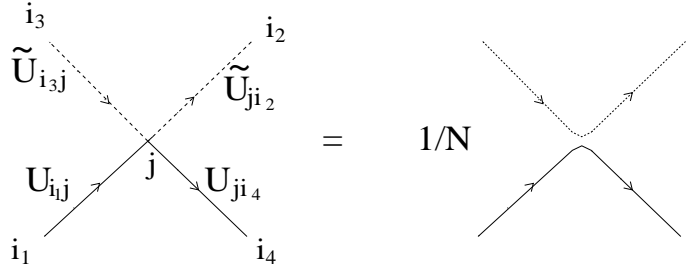}
\caption{
Parts of two Wilson lines in $S_{XY}$ that cross at site $j$.
The solid (dashed) lines represent links in $X$ ($Y$).
This crossed Wilson lines can be reconnected 
inside the average of $< ... >_Y$ 
so that links in $X$ ($Y$) form a Wilson line by themselves (see the text).
}
\label{fig:operation_x}
\end{figure}

We note that $S_{XY}[U,\td U]$ consists of Wilson loops where
parts of the loops are made of holonomy in $X$
and the remaining parts in $Y$.
For example, the loop in Fig. \ref{fig:loops_examples} (a) represents a term
\bqa
N \sum_{L_1} \JJ_{L_1} \tr [ H_{L_{1,1}} \td H_{ L_{1,2}} ]
\eqa
in $S_{XY}$, 
where $L_{1,1}$ and $L_{1,2}$ are two segments of the closed loop $L_1$
which belong to the sub-lattices $X$ and $Y$ respectively.
Hereafter, we will use the notation 
$L_a = \sum_b L_{a,b}$
to represent that $L_{a,b}$ is
the $b$-th segment of a closed loop $L_a$,
where $b$ increases along the orientation.
Accordingly, $H_{L_{a,b}}$ represents
the holonomy along the curves $L_{a,b}$.
We add tilde as $\td H_{L_{a,b}}$ 
to represent holonomies along 
curves which are in $Y$.
For all diagrams in Fig. \ref{fig:loops_examples}, 
we have \bqa
H_{L_{a,1}}  =  \prod_{ <i,j> \in L_{a,1} } U_{ij}, ~~~~~
\td H_{L_{a,2}}  =  \prod_{ <i,j> \in L_{a,2} } \td U_{ij}.
\label{open}
\eqa
These holonomies on open segments are the basic building blocks 
for the coupling between the two sub-lattices in $S_{XY}$.
However, the present form of the coupling is not very convenient for real space RG
because holonomies on open segments in each sub-lattice
are not gauge invariant. 
Therefore it is desirable to reorganize terms in $S_{XY}$ such that 
the gauge symmetry is more manifest in each sub-lattice, 
that is, $S_{XY}$ is written as products of gauge invariant
Wilson loops in each sub-lattice.
To achieve this, we first note that 
the action $S_Y[\td U]$ is invariant under a `sub-lattice gauge transformation'
defined by
\bqa
\td U_{ij} & \rightarrow & V_i^\dagger \td U_{ij} V_j, ~~~ \mbox{ for $<i,j> \in Y$ }, \nn
U_{ij} & \rightarrow & U_{ij}, ~~~ \mbox{for $<i,j> \in X$},
\eqa
where $V_i$ is $U(N)$ matrix.
Note that this is not a part of the original gauge transformation\footnote{
Even though $\td U_{ij}$ with $<i,j>\in Y$ and $U_{ik}$ with $<i,k> \in X$ 
share the same site $i$ only $\td U_{ij}$ gets transformed under the sub-lattice gauge transformation.
}. 
The full action is not invariant under this transformation because
$S_{XY}$ includes holonomies on open segments in sub-lattice $Y$.
Since the action $S_Y[\td U]$ respects the sub-lattice gauge symmetry in $Y$,
only those configurations
 where there is no open end in $Y$ survive
inside $< ... >_Y$.
Therefore, $\left< \tr [ H_{L_{1,1}} \td H_{L_{1,2}} ] \right>_Y$
for Fig. \ref{fig:loops_examples} (a) vanishes.
Segments in $Y$
may form closed loops by themselves 
as in Fig. \ref{fig:loops_examples} (b),
in which case the average does not vanish.
Non-vanishing contributions can be written
as a product of Wilson loops using the formula
\footnote{To prove this,
consider $< ... \td U_{i_3 j; \alpha \beta} \td U_{j i_2; \gamma \delta} ... >_Y$,
where $\alpha$, $\beta$, $\gamma$, $\delta$ are color indices,
while $j, i_2, i_3$ are site indices,
and $...$ represents insertions of link fields which do not include the site $j$.
The color index $\beta$ ($\gamma$) carry
fundamental (anti-fundamental) charge of U(N)
at site $j$.
Because $S_Y$ is gauge invariant,
we have
$< ... \td U_{i_3 j; \alpha \beta} \td U_{j i_2; \gamma \delta} ... >_Y
= V_{\beta^{'} \beta} V^*_{\gamma^{'} \gamma} < ... \td U_{i_3 j; \alpha \beta^{'}} \td U_{j i_2; \gamma^{'} \delta} ...  >_Y$ for any $U(N)$ matrix $V$.
The only $U(N)$ singlet tensor that can be constructed from one fundamental
and one anti-fundamental indices
is the identity.
Therefore
$< ... \td U_{i_3 j; \alpha \beta} \td U_{j i_2; \gamma \delta} ... >_Y = A_{\alpha \delta} \delta_{\beta \gamma}$
with $A_{\alpha \delta} = \frac{1}{N} < ... \td U_{i_3 j; \alpha \beta^{'}} \td U_{j i_2; \beta^{'} \delta} ... >_Y$.
From this, one obtains Eq. (\ref{formula_x}).},
\bqa
\left< \tr( A U_{i_1 j} \td U_{j i_2} \td B) \tr ( \td C \td U_{i_3 j} U_{j i_4} D ) \right>_Y
& = & \frac{1}{N}
\left< \tr( A U_{i_1 j} U_{j i_4} D) \tr ( \td C \td U_{i_3 j} \td U_{j i_2} \td B ) \right>_Y.
\label{formula_x}
\eqa
This is illustrated in Fig. \ref{fig:operation_x}.
We emphasize that this identity is valid only inside the average,
but not as an operator identity.
Using this formula, we obtain
\bqa
\left< \tr [ H_{L_{1,1}} \td H_{ L_{1,2}} ] \right>_Y
&=& \frac{1}{N} \left< \tr [ H_{L_{1,1}}] \tr[ \td H_{ L_{1,2}} ] \right>_Y \nn
&=& \frac{1}{N} \left< W_{L_{1,1}} \td W_{ L_{1,2}} \right>_Y
\eqa
for Fig. \ref{fig:loops_examples} (b).
Generally, only those diagrams where
all open segments in sub-lattice $Y$  form
closed loops do not vanish.
For Fig. \ref{fig:loops_examples} (c) and (d),
one obtains
\bqa
\left< \tr [ H_{L_{1,1}} \td H_{ L_{1,2}} ] \tr [ H_{L_{2,1}} \td H_{L_{2,2}} ]\right>_Y
&=& \frac{1}{N^2} \left< W_{L_{1,1}+L_{2,1}} \td W_{ L_{1,2} + L_{2,2}} \right>_Y, \nn
\left< \tr [ H_{L_{1,1}} \td H_{ L_{1,2}} ] \tr [ H_{L_{2,1}} \td H_{L_{2,2}} ]
\tr [ H_{L_{3,1}} \td H_{L_{3,2}} ]
\right>_Y
&=& \frac{1}{N^3} \left< W_{L_{1,1}+L_{2,1}+L_{3,1}} 
\td W_{ L_{1,2} + L_{3,2}+ L_{2,2}} \right>_Y. 
\eqa
Note that if segments in sub-lattice $Y$ form loops,
segments in sub-lattice $X$ automatically
form loops.
Therefore all non-vanishing terms in $\left< e^{- S_{XY}[U, \td U]} \right>_Y$
can be expressed as a polynomial of Wilson loops in $X$ and $Y$,
\bqa
\left< e^{- S_{XY}[U, \td U]} \right>_Y
&=& \left<
1 +
\sum_{n=1}^\infty \sum_{m=1}^\infty
\sum_{\CCC \in X} \sum_{\CCCT \in Y}
F_{\CCC;\CCCT}[\JJ]
 \prod_{i=1}^n W_{C_i}
\prod_{k=1}^m W_{\td C_k}
\right>_Y, \nn
\eqa
where $F_{\CCC;\CCCT}[\JJ]$ depends only on $\JJ$, $\CCC$ and $\CCCT$,
but not on $U_{ij}$, $\td U_{ij}$\footnote{When 
more than two Wilson loops cross at a point, 
one needs more general formula than Eq. (\ref{formula_x}). 
However, all non-vanishing contributions can be always 
written as products of Wilson loops.}.
This polynomial for $W_C$ and $W_{\td C}$ can be exponentiated as
\bqa
\left< e^{- S_{XY}[U, \td U]} \right>_Y
= \left< e^{-S^{'}[W,\td W]} \right>_Y,
\eqa
where
\bqa
S^{'}[W,\td W] &=&
- \sum_{n=1}^\infty \sum_{m=1}^\infty
N^{2-(m+n)} \sum_{\CCC \in X} \sum_{\CCCT \in Y}
h_{\CCC;\CCCT}[\JJ]
 \prod_{i=1}^n W_{C_i}
\prod_{k=1}^m W_{\td C_k}. \nn
\eqa
Here $N^{2-(m+n)} h_{\CCC;\CCCT}[\JJ]$ 
is the cumulant of $F_{\CCC;\CCCT}[\JJ]$.
Since $\ln Z  \sim O(N^2)$,
$h_{\CCC;\CCCT}[\JJ]$ is $O(1)$ in the large N limit
\footnote{This is why we deliberately singled out the 
factor of $N^{2-(m+n)}$ out of $h_{\CCC;\CCCT}[\JJ]$.}.
$h_{\CCC;\CCCT}[\JJ]$ can be computed 
perturbatively in $\JJ_\CCC$ and $1/N$\footnote{This corresponds to a strong 't Hooft coupling expansion in the large $N$ limit.},
\bqa
h_{\{C_1\}; \{\td C_1\}}[\JJ] & = & 
\sum_{L_1} \JJ_{L_1} \delta_{C_1+\td C_1,L_1}
+ \JJ_{\{C_1,\td C_1\}}  \nn
&  & + \frac{1}{2} \sum_{L_1, L_2} \JJ_{L_1} \JJ_{L_2}
\delta_{L_{1,1}+L_{2,1},C_1} \delta_{L_{1,2}+L_{2,2},\td C_1}  \nn
&  & + \frac{1}{6} \sum_{L_1, L_2, L_3} \JJ_{L_1} \JJ_{L_2} \JJ_{L_3}
\delta_{L_{1,1}+L_{2,1}+L_{3,1},C_1} \delta_{L_{1,2}+L_{3,2}+L_{2,2},\td C_1}\nn
&& + O(\JJ^4,1/N), \label{h1} \\
h_{\{C_1,C_2\}; \{\td C_1\}}[\JJ] & = & \sum_{L_1} \JJ_{L_1}
\delta_{L_{1,1},C_1}
\delta_{L_{1,2}+L_{1,4},\td C_1}
\delta_{L_{1,3},C_2} \nn
&& + \sum_{L_1,L_2}
 \JJ_{\{L_1, L_2\}}
\delta_{L_{1,1},\td C_1}
\delta_{L_{1,2},C_1}
\delta_{L_2,C_2} \nn
&& + \JJ_{\{C_1,C_2,\td C_1\}}  + O(\JJ^2, 1/N), 
\label{h2}
\\
h_{\{C_1\}; \{\td C_1, \td C_2 \}}[\JJ] & = & \sum_{L_1} \JJ_{L_1}
\delta_{L_{1,1},\td C_1}
\delta_{L_{1,2}+L_{1,4}, C_1}
\delta_{L_{1,3},\td C_2} \nn
&& + \sum_{L_1,L_2}
 \JJ_{\{L_1, L_2\}}
\delta_{L_{1,1}, C_1}
\delta_{L_{1,2},\td C_1}
\delta_{L_2,\td C_2} \nn
&& + \JJ_{\{C_1,\td C_1, \td C_2\}}  + O(\JJ^2, 1/N),
\label{h3}
\eqa
and so on.
Here 
$\delta_{C_1, C_2}$ is a Kronecker delta function
in the space of loops.
It is easy to understand physical meaning of each term.
The first term in Eq. (\ref{h1}) describes a loop $L_1$
which has a self-intersecting point as in Fig. \ref{fig:loops_examples} (b)
decomposes into two loops $C_1$ and $\td C_1$ 
which are in $X$ and $Y$, respectively.
The second term describes a two-loop state
where one is included in $X$ and 
the other in $Y$.
The third and fourth terms describe
the process where two and three loops 
join to form a loop $C_1$ in $X$
and a loop $\td C_1$ in $Y$ 
as in Fig. \ref{fig:loops_examples} (c) and (d), respectively. 
The first term in Eq. (\ref{h2}) describes
a loop $L_1$ decomposing into three loops,
two of which ($C_1$, $C_2$) are in $X$ and 
one ($\td C_1$), in $Y$
through two self intersecting points.
The remaining terms can be understood similarly.

We call those Wilson loops that contribute to
$h_{\CCC;\CCCT}[\JJ]$ 
`connected Wilson loops'.
There are two kinds of them.
Connected Wilson loops of the first kind are those
that touch each other in space at crossing points
where links in $X$ and $Y$ meet.
For example, the Wilson loops $C_1$ and $\td C_1$
in the first term of Eq. (\ref{h1}) are connected at a 
crossing point as in Fig. \ref{fig:loops_examples} (b). 
The second kind includes those Wilson loops that
are physically separated but the separated loops 
originate from multi-trace couplings, such as 
the second term in Eq. (\ref{h1}), where
$C_1$ and $\td C_1$ are in general separated in space 
but they are `connected' through  $\JJ_{C_1, \td C_1}$.

Now the partition function can be written as
\bqa
Z[\JJ] &=& \int_X d U
\int_Y d \td U
e^{-S_Y[\td U] + \sum_{n=1}^\infty  \sum_{\CCC \in X} N^{2-n}
( \JJ_\CCC + f_\CCC[\td W]) \prod_{i=1}^n W_{C_i}},
\eqa
where
\bqa
f_\CCC [\td W] = \sum_{m=1}^\infty
N^{-m} \sum_{\CCCT \in Y}
h_{\CCC;\CCCT}[\JJ]
\prod_{k=1}^m W_{\td C_k}.
\eqa
This theory can be viewed as a theory defined on 
the lattice $X$ whose sources
$\JJ_\CCC + f_\CCC[\td W]$
are dynamical, 
where fluctuations of the sources
are provided by the dynamical degrees of freedom
defined on $Y$.
Since  $W_{\td C_k} \sim O(N)$,
the dynamical sources become classical in the large $N$ limit.
Therefore it is useful to introduce collective fields 
for the source fields.
We decompose $\td W$ and $W$ by introducing the Hubbard Stratonovich fields\cite{SLEE10},
\bqa
Z[\JJ]
&=& \int d U d \td U dJ dP
 e^{-\left( S_Y[\td U] + S^{''}[J,P,W,\td W] \right) },
\eqa
where $dJ dP \equiv \prod_{n=1}^\infty \prod_{\CCC \in X} dJ_\CCC dP_\CCC$ and
\bqa
S^{''}[J,P,W,\td W] & = &
i N^2 \sum_{n=1}^\infty  \sum_{\CCC \in X}
P_\CCC \left( J_\CCC - \JJ_\CCC - f_\CCC[\td W] \right) \nn
&&  - \sum_{n=1}^\infty  \sum_{\CCC \in X} N^{2-n} J_\CCC \prod_{i=1}^n W_{C_i}.
\eqa
$J_\CCC$'s are fluctuating sources for Wilson loops on $X$,
and $P_\CCC = P_{\{ \bar C_1,...,\bar C_n \}}^*$ is a complex Lagrangian multiplier
which imposes the constraint
$J_\CCC =  \JJ_\CCC + f_\CCC[\td W]$
and its complex conjugate.
With this normalization of $P_\CCC$, 
the equation of motion for $J_\CCC$ implies
$i < P_\CCC > = N^{-n} \left< \prod_{i=1}^n W_{C_i} \right> \sim O(1)$.
Physically, $P_\CCC$ describes fluctuations of the Wilson loop operators.
The dynamical action for $P_\CCC$ is generated
once $\td U$ is integrated over,
\bqa
Z[\JJ]
&=& \int d U dJ dP
 e^{-\left( S_D[\JJ,J,P] + S_X[U;J] \right) },
\eqa
where
\bqa
S_D[\JJ,J,P] & = &
i N^2 \sum_{n=1}^\infty  \sum_{\CCC \in X}
P_\CCC ( J_\CCC - \JJ_\CCC )
+ G[\JJ,P], \nn
G[\JJ,P] & = & -\ln \left< e^{ i N^2 \sum_{n=1}^\infty  \sum_{\CCC \in X}
P_\CCC f_\CCC[\td W] } \right>_Y, \nn
S_X[U;J] &=& - \sum_{n=1}^\infty  \sum_{\CCC \in X} N^{2-n} J_\CCC \prod_{i=1}^n W_{C_i}.
\eqa

\begin{figure}[h!]
\centering
        \includegraphics[height=5cm,width=12cm]{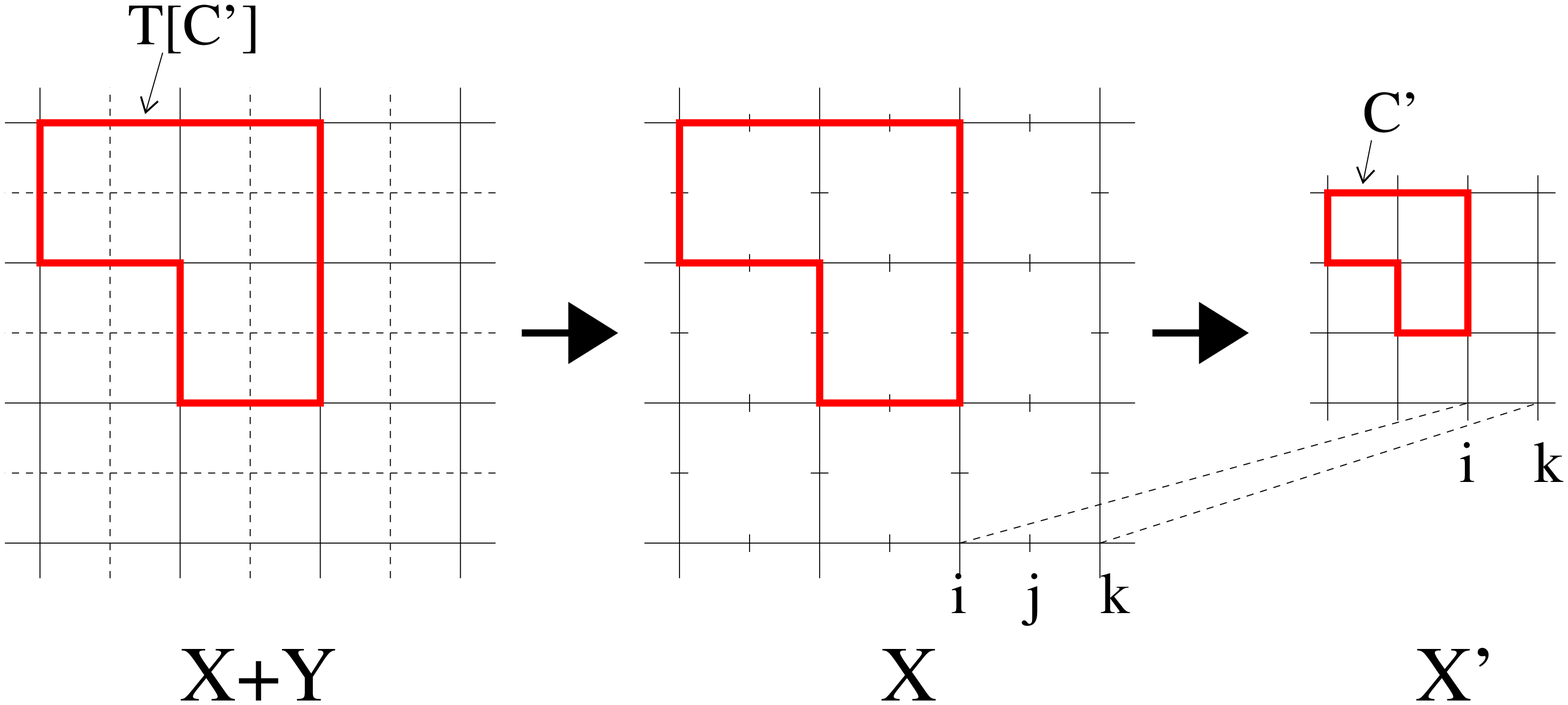}
\caption{
Procedure of coarse graining.
First, the link variables in $Y$ are integrated out.
Then, the remaining lattice $X$ is rescaled 
into a coarse grained lattice $X^{'}$ so that
every two consecutive links in $X$ without a branch
get merged into one in $X^{'}$.
}
\label{fig:rescaled_lattice}
\end{figure}

In order to repeat the coarse graining procedure,
it is convenient to map $X$ into the same form 
as the original lattice.
For this, we merge every two links in $X$ 
into one link in a new lattice $X^{'}$ 
as is shown in Fig. \ref{fig:rescaled_lattice}.
Accordingly, two holonomies on merged links $<i,j>, <j,k> \in X$ 
are combined to produce one holonomy as
$U_{ik}^{'} = U_{ij} U_{jk}$
which is defined
on the coarse grained lattice labeled by $<i,k> \in X^{'}$.
Since the Wilson loops on $X$ depend only on $U^{'}$,
the partition function becomes
\bqa
Z[\JJ]
&=& \int d U^{'} dJ^{'} dP^{'}
 e^{-\left( S_D^{'}[\JJ,J,P] + S_{X^{'}}[U^{'};J] \right) }
\eqa
upto an unimportant multiplicative constant,
where $dJ^{'} dP^{'} \equiv \prod_{n=1}^\infty \prod_{\CCCP \in X^{'}} dJ_\CCCP dP_\CCCP$ and
\bqa
S_D^{'}[\JJ,J,P] & = &
i N^2 \sum_{n=1}^\infty  \sum_{\CCCP \in X^{'}}
P_\CCCP ( J_\CCCP - \JJ_\DCCCP )
+ G^{'}[\JJ,P], 
\label{SD}
\\
G^{'}[\JJ,P] & = & -\ln \left< e^{ i N^2 \sum_{n=1}^\infty  \sum_{\CCCP \in X^{'}}
P_\CCCP f_\DCCCP[\td W] } \right>_Y, 
\label{GP} \\
S_{X^{'}}[U^{'};J] &=& - \sum_{n=1}^\infty  \sum_{\CCCP \in X^{'}} N^{2-n} J_\CCCP \prod_{i=1}^n W_{C_i^{'}}.
\label{SP}
\eqa
Here $C_i^{'}$'s are loops on $X^{'}$.
$T$ represents a dilatation map
which takes a loop in $X^{'}$ 
to the original one before rescaling in $X$ 
as is illustrated in Fig. \ref{fig:rescaled_lattice}.
This is necessary because $J_\CCCP$ and $P_\CCCP$ 
are defined on $X^{'}$
while $\JJ_\CCC$ and $f_\CCC[\td W]$ are defined 
on $X \subset X+Y$.

\begin{figure}[h!]
\centering
        \includegraphics[height=2.5cm,width=7cm]{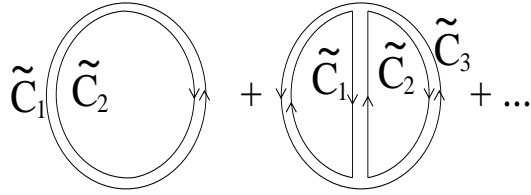}
\caption{
First two leading order planar diagrams for $G^{'}[\JJ,P]$.
These loops are defined on $Y$ and each loop is associated
with $r_\CCCT$.
}
\label{fig:GJP}
\end{figure}

This is a theory defined on the coarse grained lattice $X^{'}$
with dynamical coupling fields $J_\CCCP$, $P_\CCCP$ 
with the action $S_D^{'}$.
The second term in Eq. (\ref{SD}) is given by 
\bqa
G^{'}[\JJ,P] & = & -\ln \int d \td U e^{-S_Y^{'}[\td U]},
\eqa
where
\bqa
S_Y^{'}[\td U] = - \sum_{m=1}^\infty  \sum_{\CCCT \in Y} N^{2-m}
r_{ \CCCT }
\prod_{i=1}^m W_{\td C_i}
\eqa
with
\bqa
r_\CCCT =
\JJ_\CCCT +
i \sum_{n=1}^\infty
\sum_{\CCCP \in X^{'}}
P_\CCCP h_{\DCCCP;\CCCT}[\JJ].
\label{eq:r}
\eqa
Therefore $G^{'}[\JJ,P]$ is given by the effective potential
for the gauge theory defined on the sub-lattice $Y$
with the set of couplings, $r_\CCCT$.
In the large $N$ limit,
planar diagrams give the leading contribution
and $G^{'}[\JJ,P]$ is $O(N^2)$.
For large 't Hooft couplings ($\JJ << 1$),
it is natural to compute $G^{'}[\JJ,P]$ as a
power series of $r_\CCCT$ in the real space\cite{BARS},
\bqa
G^{'}[\JJ,P] = - N^2 \left[ \frac{1}{2} \sum_{\td C_1, \td C_2 \in Y}
r_{\td C_1}
r_{\td C_2}
\delta_{\td C_1 + \td C_2, 0}
+ \frac{1}{6} \sum_{\td C_1, \td C_2, \td C_3 \in Y}
r_{\td C_1}
r_{\td C_2}
r_{\td C_3}
\delta_{\td C_1 + \td C_2 + \td C_3, 0} + ... \right].
\label{eq:G2}
\eqa
This can be easily checked by using 
$\int d \td U_{ij} ~ \td U_{ij;\alpha \beta} \td U_{ji; \gamma \delta}
= \frac{1}{N} \delta_{\alpha, \delta} \delta_{\beta, \gamma} 
[ \int d \td U_{ij} ]$.
Due to the Kronecker delta function,
the first term survives only when $\td C_1 = \bar{ \td{ C_2}}$.
Similarly, the second term is nonzero 
only when the three oriented loops form a connected double-line graph
with no unpaired single line.
These are illustrated in Fig. \ref{fig:GJP}.
General contributions to $G^{'}[\JJ,P]$
are given by multiple oriented surfaces made of double lines,
where each face in double line graphs
is associated with $r_\CCCT$.
For example,
$-N^2
\prod_{k} r_{\td C_{k}}
\delta_{\sum_k \td C_{k}, 0}
$
and
$-N^2 r_{\td C_1, \td C_2}
\prod_{k} r_{\td C_{k}^{'}}
\prod_{l} r_{\td C_{l}^{''}}
\delta_{\td C_1 + \sum_k \td C_{k}^{'}, 0}
\delta_{\td C_2 + \sum_l \td C_{l}^{''}, 0}
$
are contributions from one surface and two surfaces, respectively.
Note that $G^{'}[\JJ,P]$ is non-linear in $P_\CCC$, 
and fluctuations of $P_\CCC$ no longer impose
a strict delta function for $J_\CCC$ :
they become dynamical fields.

\begin{figure}[h!]
\centering
        \includegraphics[height=7cm,width=8cm]{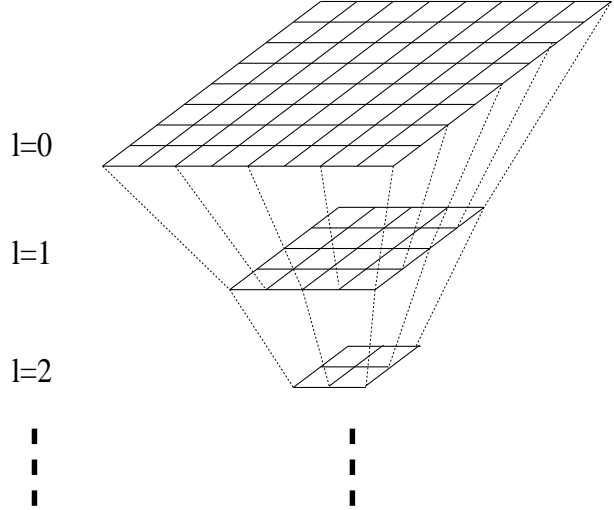}
\caption{
The `lattice AdS' space. 
The sequence of $D$-dimensional coarse grained lattices form 
a $(D+1)$-dimensional space on which the lattice loop field theory
is defined.
}
\label{fig:lattice_AdS}
\end{figure}

This completes one step of our RG procedure.
If we repeatedly apply this cycle to the theory on $X^{'}$,
and then to the coarse grained lattice of $X^{'}$ 
and so on, 
we can write the partition function as
a integration over $J^{(l)}$'s and $P^{(l)}$'s\cite{SLEE10},
\bqa
Z[\JJ] &=& \int \prod_{l=1}^{\infty} \left[ dJ^{(l)} dP^{(l)} \right]
 e^{- S_{LLFT} }, \nn
S_{LLFT} & = & \sum_{l=0}^\infty S_D^{'}[J^{(l)},J^{(l+1)},P^{(l+1)}],
\eqa
where $J^{(l)}$ and $P^{(l)}$ are Hubbard-Stratonovich fields
introduced at the $l$-th step of coarse graining, and
$J^{(0)}$ is fixed by the microscopic couplings,
\bqa
J_\CCC^{(0)} = \JJ_\CCC.
\label{eq:BC}
\eqa
Here $l$ plays the role of a discrete coordinate
for the new $(D+1)$-th dimension that corresponds
to the length scale in RG;
fields at small (large) $l$ describe
UV (IR) physics.
This $(D+1)$-dimensional theory is a holographic theory for the
$D$-dimensional $U(N)$ lattice gauge theory.
We will call it {\it lattice loop field theory} (LLFT).

Since it is difficult to write down the full theory
in a compact form,
let us try to understand some general features of the theory
from the first few leading terms of the action in the strong coupling expansion ($J << 1$).
By plugging  Eqs. (\ref{h1})-(\ref{h3})
into Eqs. (\ref{eq:r}) and (\ref{eq:G2}),
one obtains,
\bqa
S_{LLFT} & = & 
 N^2 \sum_{l=0}^\infty \left[ i \sum_{n=1}^\infty  \sum_\CCC
P_\CCC^\HKP ( J_\CCC^\HKP - J_\DCCC^\HK ) + {\cal H}[J^\HK,P^\HKP] \right], \nn
\label{LFT}
\eqa
where the `Hamiltonian' ${\cal H}$
(the reason for this naming will become clear shortly) 
 is given by
\bqa
&& {\cal H }[J,P]  =  
- \frac{1}{2} \Big\{
J_{\td C}
+ i P_{C_1} 
\left( 
J_{L_1} \delta_{L_1,T[C_1]+\td C}
+ J_{\{T[C_1],\td C\}}
+ \frac{1}{2} J_{L_1} J_{L_2} \delta_{L_{1,1}+L_{2,1},T[C_1]} \delta_{L_{1,2}+L_{2,2},\td C} \right) \nn
&& ~~~~~~ + i P_{\{C_1,C_2\}}
\Big(
J_{L_1} \delta_{L_{1,1},T[C_1]}
\delta_{L_{1,2}+L_{1,4},\td C}
\delta_{L_{1,3},T[C_2]}
+ J_{\{L_1,L_2\}} \delta_{L_{1,1},\td C}
\delta_{L_{1,2},T[C_1]}
\delta_{L_2,T[C_2]} \nn
&& ~~~~~~ + J_{\{T[C_1],T[C_2],\td C\}}
\Big) + ... 
\Big\} \times \nn
&& ~~~ \Big\{
J_{ \bar{ \td C} }
+ i P_{C_3}
\left( 
J_{L_3} \delta_{L_3,T[C_3]+\bar{\td C}}
+ J_{\{T[C_3],\bar{\td C}\}}
+ \frac{1}{2} J_{L_3} J_{L_4} \delta_{L_{3,1}+L_{4,1},T[C_3]} \delta_{L_{3,2}+L_{4,2},\bar{\td C}} \right) \nn
&& ~~~~~~ + i P_{\{C_3,C_4\}}
\Big(
J_{L_3} \delta_{L_{3,1},T[C_3]}
\delta_{L_{3,2}+L_{3,4},\bar{\td C}}
\delta_{L_{3,3},T[C_4]}
+ J_{\{L_3,L_4\}} \delta_{L_{3,1},\bar{\td C}}
\delta_{L_{3,2},T[C_3]}
\delta_{L_4,T[C_4]} \nn
&& ~~~~~~ + J_{\{T[C_3],T[C_4],\bar{\td C}\}}
\Big) + ...
\Big\} 
 + ...,
\label{H}
\eqa
where $...$ include terms that involve
fields associated with multi-loop states 
and higher order terms in $J, P, 1/N$. 
Here indices 
$L_i$, $C_i$ and $\td C$ are understood to be summed
over loops in $X+Y$, $X^{'}$ and $Y$, respectively.
If $\td C$ is a self-retracting loop, we set $J_{\td C} = 1$.
LLFT is defined on
the $(D+1)$-dimensional lattice
shown in Fig. \ref{fig:lattice_AdS}.
This lattice may be viewed as a
discrete version of the anti-de Sitter (AdS) space
where the continuous isometry 
$l = l^{'} + \alpha$, $x^\mu = e^{\alpha} x^{\mu '}$
of the metric $ds^2 = dl^2 + e^{-2l} \sum_{\mu=1}^D dx^{\mu 2}$
is replaced by a discrete scale invariance,
$l = l^{'} + 1$, $x_i = 2 x_i^{'}$,
if one assumes that all nearest neighbor bonds along the D-dimensional directions
have the same physical length.
However, we emphasize that this assumption is not generally true, 
and the metric is determined dynamically from the equation of motion for the loop fields.
The true AdS space will emerge only if the solution
respects the scale invariance.
In theories which are not conformal, such as the present pure Yang-Mills theory, 
we expect that the infrared geometry will be effectively cut-off
by the mass gap.
We will defer this dynamical issue to a future study, 
and focus on the general structure of the theory in the following.
We note that this kind of `discrete AdS' space 
has been the natural setting for the real
space renormalization group approach 
and the multi-scale entanglement renormalization Ansatz (MERA)\cite{VIDAL_MERA,GU,SWINGLE09}.
The degrees of freedom of LLFT
are fields of loops $J_\CCC^{(l)}$ and $P_\CCC^{(l)}$.

It is convenient to interpret $l$ as a discrete (imaginary) `time'.
Then we can identify the first term in Eq. (\ref{LFT})
as the Berry phase term which
dictates that $J_\CCC^{(l)}$ and $P_\CCC^{(l)}$ are conjugate to each other :
loop fields as operators would
satisfy the commutation relation $[J_\CCC, P_\CCC] = i/N^2$
if time was continuous.
The remaining term ${\cal H}$ in Eq. (\ref{LFT})
is the `Hamiltonian' that governs the
evolution of the loop fields along the discrete time\footnote{
However, the analogy with Hamiltonian is not perfect because
${\cal H}$ is not Hermitian.}.
Formally, $J_\CCC^{(l)}$ ($i P_\CCC^{(l)}$) can be
viewed as the path integration representation
of an operator that annihilates (creates)
a set of loops $\CCC$ at time $l$, and
$J_{{\bar C_1,..,\bar C_2}}^\HK$ 
($i P_{{\bar C_1,..,\bar C_2}}^\HK$)
is associated with an annihilation (creation) operator of `anti-loops'.
It is noted that $J_\CCC$ and $i P_\CCC$
are not the usual annihilation and creation
operators because they are not Hermitian conjugate to each other.
In the basis given by
\bqa
a_\CCC & = & \frac{N}{\sqrt{2}} (J_\CCC - i P_\CCCB) , \nn
b_\CCC & = & \frac{N}{\sqrt{2}} (J_\CCCB - i P_\CCC), 
\eqa
$a_\CCC$ and $a_\CCC^\dagger$ 
($b_\CCC$ and $b_\CCC^\dagger$) 
become the standard annihilation and creation operators
of loops $\CCC$ (anti-loops $\CCCB$) respectively.
However, we will use the basis of $J_\CCC$ and $P_\CCC$ in the following because it has two merits.
First, the Hamiltonian becomes particularly simple in this basis
because ${\cal H}$ contains only those $P_\CCC$ with $C_i \in X^{'}$
: loops can be created only on those links that survive 
coarse graining at each step of RG.
Second, the equations of motion for $J_\CCC$ and $J_\CCCB$
implies $i <P_\CCC> = [ i <P_\CCCB> ]^*$.
Therefore it is convenient to absorb $i$ into $P_\CCC$ to
define ${\cal P}_\CCC = i P_\CCC$.
At the saddle point, the Hamiltonian becomes Hermitian 
and the Berry phase term for ${\cal P}_\CCC$ 
and $J_\CCC$ becomes that of the standard 
creation and annihilation operators upto 
the factor of $N^2$.
Note that the expectation value of creation operator 
is not in general complex conjugate of that of 
annihilation operator,
i.e. $ <{\cal P}_\CCC> \neq < J_\CCC>^*$,
when the saddle point is `time'-dependent.

\begin{figure}[h!]
\centering
        \includegraphics[height=6cm,width=10cm]{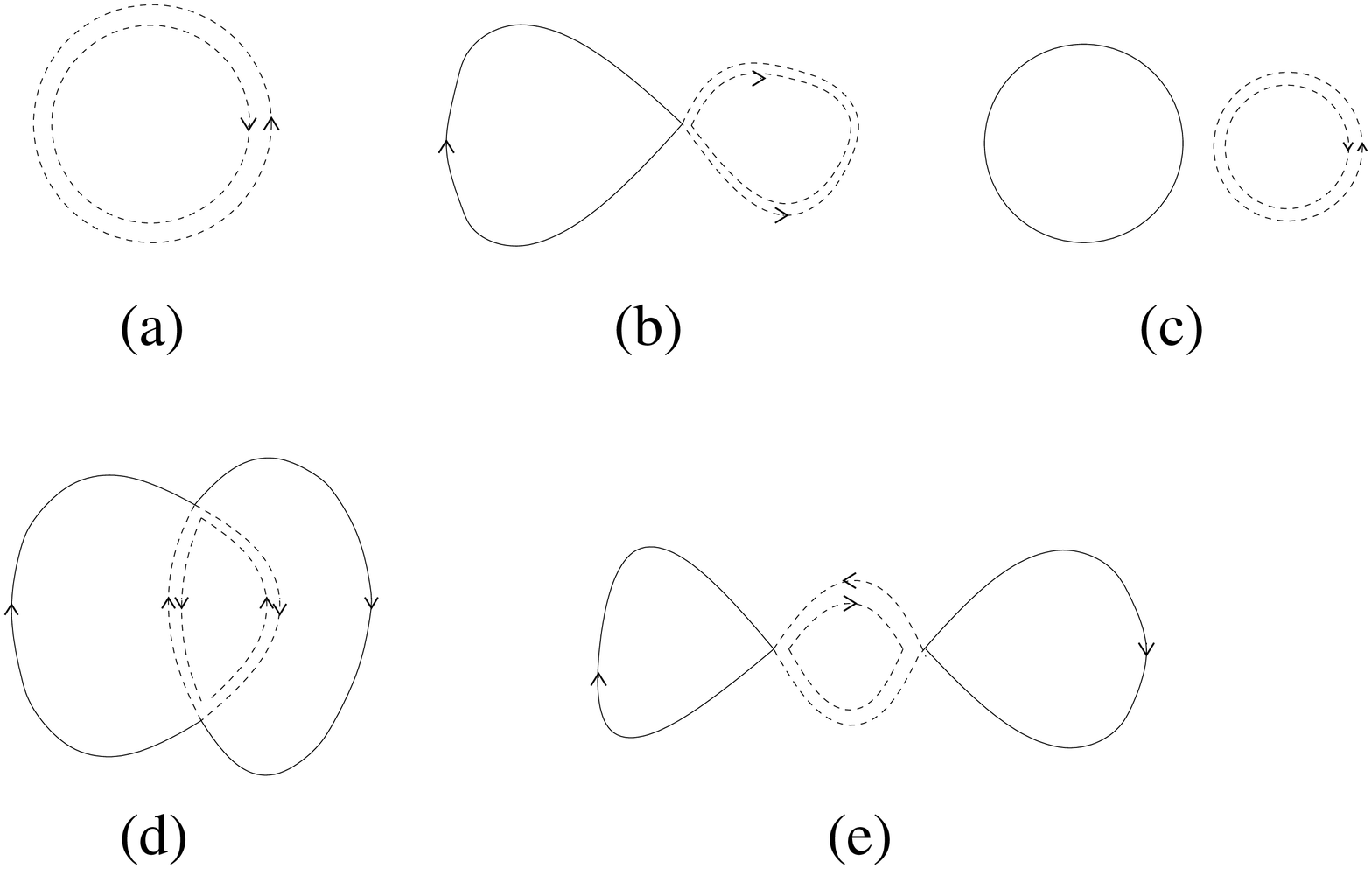}
\caption{
Diagrams that contribute to the Hamiltonian.
}
\label{fig:loop_interaction}
\end{figure}

Now let us take a closer look at the Hamiltonian 
to understand the physical meaning of each term.
The quadratic term $J_{\td C} J_{\bar{ \td C}}$ describes
a process where 
a loop $\td C$ and its anti-loop $\bar{ \td C}$
are pair-annihilated (Fig. \ref{fig:loop_interaction} (a)).
Higher order terms describe fluctuations and
joining/splitting processes of loops :
$i P_{C_1} J_{L_1} \delta_{L_1,T[C_1]+\td C} J_{\bar{ \td C}}$ 
describes a process where
a loop $L_1$ with a self intersection becomes 
a smaller loop by combining with a loop $\bar{\td C}$ (Fig. \ref{fig:loop_interaction} (b));
$i P_{C_1} J_{\{T[C_1], \td C\}} J_{\bar{\td C}}$
describes one of the loops in a two-loop state
disappearing into vacuum with its anti-loop (Fig. \ref{fig:loop_interaction} (c));
$i P_{C_1} J_{L_1} J_{L_2} \delta_{L_{1,1}+L_{2,1},T[C_1]} \delta_{L_{1,2}+L_{2,2},\td C} J_{\bar{ \td C}}$
describes two loops merging into one loop 
with a help of an anti-loop which eliminates
parts of the two loops(Fig. \ref{fig:loop_interaction} (d));
$i P_{\{C_1,C_2\}}  
J_{L_1} \delta_{L_{1,1},T[C_1]}
\delta_{L_{1,2}+L_{1,4}, \td C}
\delta_{L_{1,3},T[C_2]} J_{\bar{\td C}}$
describes a process where
one loop gets split into two(Fig. \ref{fig:loop_interaction} (e)).

\begin{figure}[h!]
\centering
        \includegraphics[height=4cm,width=12cm]{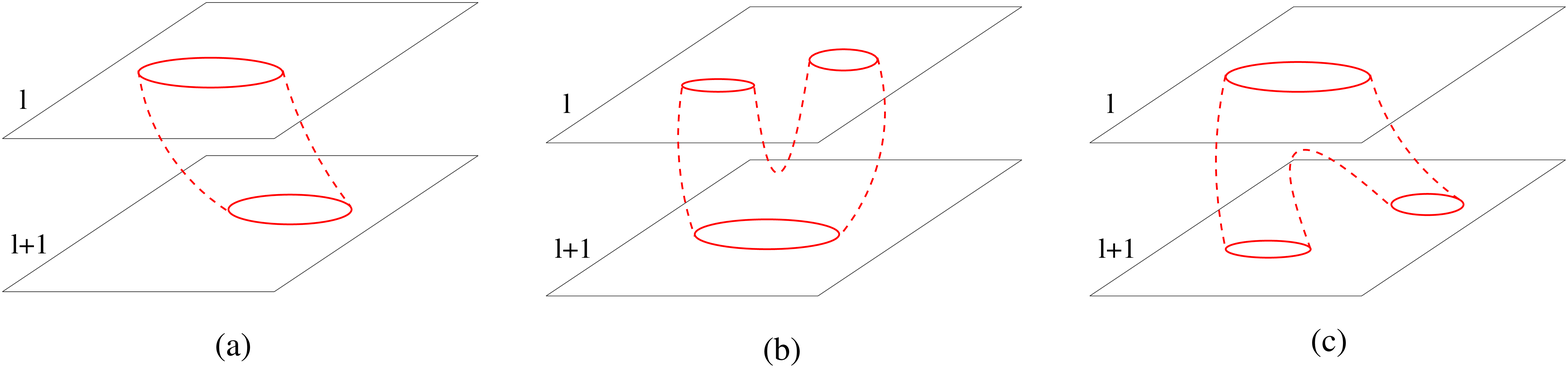}
\caption{
The quadratic and cubic terms for the
propagating loop fields $j_C$, $p_C$
in the Hamiltonian.
The quadratic term (a) describes fluctuation
and propagation of loops, 
and the cubic terms (b), (c) describe
joining and splitting processes of loops.
}
\label{fig:perturbative_loop_interaction}
\end{figure}

Usually, it is expected that
a quadratic kinetic energy term describes
propagation of loops in space and time.
Actually there is no such term in the Hamiltonian.
Instead, fluctuations of loop arise only through cubic
and higher order terms in $J$ and $P$.
Here we have to remember that 
$J_\CCC^\HK$ and $P_\CCC^\HK$ have
nonzero expectation value in the vacuum
because of the boundary condition
$\JJ_\CCC^{(0)} \neq 0$. 
Therefore one has to identify 
$j_\CCC = J_\CCC - <J_\CCC>$ and 
$p_\CCC = P_\CCC - <P_\CCC>$ 
as the propagating loop fields.
For these fields that describe small fluctuations 
of loop fields, 
the Hamiltonian includes
the quadratic kinetic energy 
which describes fluctuations
and propagation of loops in spacetime,
\bqa
\sum_{C_1,C_2} 
t_{C_1, C_2} 
p_{C_1}^\HKP 
j_{C_2}^\HK,
\label{ke}
\eqa
where $t_{C_1,C_2}$ is 
`hopping' probability amplitude 
for a loop $C_2$ at time $l$ to change into a loop $C_1$ 
at time $l+1$.
The Hamiltonian also contains
interactions between propagating loop fields.
For example, the cubic interaction terms,
\bqa
\sum_{C_1,C_2,C_3} V_{C_1,C_2,C_3} 
p_{C_1}^\HKP j_{C_2}^\HK j_{C_3}^\HK, \nn
\sum_{C_1,C_2,C_3} V^{'}_{C_1,C_2,C_3} 
p_{C_1}^\HKP p_{C_2}^\HKP j_{C_3}^\HK
\eqa
describe the processes where two loops join into one loop
and one loop splits into two loops, respectively.
These are illustrated in Fig. \ref{fig:perturbative_loop_interaction}.
In general, loops involved in hopping and interaction 
can be far from each other in space.
However, such non-local terms will be exponentially suppressed
if $<J_\CCC>$ is small, and decreases exponentially
as the size of loops $\CCC$ increases or
as the separation between the loops in multi-trace couplings
increases.
This is because only connected diagrams contribute to the Hamiltonian.
For example, in order for a loop to hop by a large distance,
there must be a large number of loops that connect
the initial and final states, 
or there must be a multi-trace coupling 
that connect them.
This suggests that LLFT 
intrinsically has non-local elements, 
but a sense of locality emerges if vacuum
satisfies certain conditions.
In a sense, locality is a property 
of the vacuum rather than 
the theory itself.
We will discuss more about the condition
for locality in the next section.

\begin{figure}[h!]
\centering
        \includegraphics[height=4cm,width=10cm]{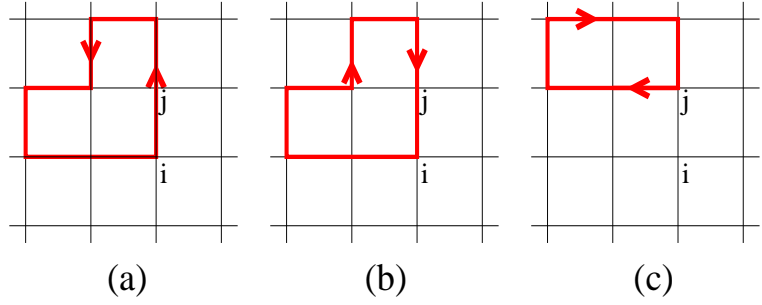}
\caption{
Three loops that contribute charge 
(a) $+1$,
(b) $-1$,
(c) $0$
to $Q^{ij}$. 
}
\label{fig:Q}
\end{figure}

Is there a symmetry underlying this action ?
The action $S_{LLFT}$ in Eq. (\ref{LFT}) 
has infinitely many conserved charges.
Namely, one can define one U(1) charge $Q^{ij} = -Q^{ji}$
for every directed link $<i,j>$ so that
\bqa
\left[ Q^{ij}, J_\CCC^\HK \right] & = &  
Y^{ij}_{\DNCCC} J_\CCC^\HK, \nn
\left[ Q^{ij}, P_\CCC^\HK \right] & = & 
-Y^{ij}_{\DNCCC} P_\CCC^\HK, 
\eqa
where $<i,j>$ is a link on the original lattice
(the boundary of the discrete AdS space),
$T^l[C_i]$ maps a loop $C_i$
on the $l$-th coarse grained lattice
to a loop on the original lattice 
by magnifying the loop by $2^l$ times 
( $T^1 =T$, $T^2 = T \circ T$, ...),
and $Y_{\CCC}^{ij}$ is the total number of times 
that loops $\CCC$ pass through the link $<i,j>$.
If a loop pass the link from $i$ to $j$ (from $j$ to $i$), 
it contribute $+1$ ($-1$) to $Y_{\CCC}^{ij}$
as is illustrated in Fig. \ref{fig:Q}.
In general, any term that respects these symmetries
arises in $S_{LLFT}$.
For example, there is a term 
that describes a multi-loop scattering process,
\bqa
\left[ \prod_{a=1}^n P_{C_a}^\HKP \right]
\left[ \prod_{b=1}^m J_{C_b^{'}}^\HK \right]
\delta_{\sum_a T[C_a], \sum_b C_b^{'} },
\eqa
where $m$ incoming loops become 
$n$ outgoing loops.
Here the delta function imposes
the charge conservation.
%

Although the action $S_{LLFT}$
is invariant under the local symmetries,
the infinite set of symmetries is broken 
by the boundary condition $J_\CCC^{(0)} \neq 0$.
In general, there remain only $D$ unbroken global symmetries
generated by,
\bqa
Q^\mu & = & \sum_i Q^{i i+\mu},
\eqa
where $\mu=1,2,...,D$.
Conservation of $Q^\mu$ implies that
if a loop has a certain number of links
along $+\mu$ direction, it should
have the same number of links along $-\mu$
to form a closed loop.
The conservation of $Q^\mu$ means that 
there are only closed loops.
This conserved charge is a lattice version
of the charge carried by world sheet 
of fundamental string
which is coupled to the NS-NS two form gauge 
field in the critical string theory.

\section{Classicality and Locality}

The prefactor $N^2$ in $S_{LLFT}$ 
can be identified as the inverse of `Planck constant'
that controls quantum fluctuations of $J_\CCC$ and $P_\CCC$.
Therefore the theory becomes classical in the large $N$ limit.
The saddle point occurs along the imaginary axis
for $P_\CCC = -i {\cal P}_\CCC$.
The equation of motion takes the form of the Hamilton equation\cite{VERLINDE} in discrete time,
\bqa
J_\CCC^\HKP - J_\DCCC^\HK & = & 
-\frac{\partial {\cal H}[J^\HK,P^\HKP] }{\partial {\cal P}_\CCC^\HKP }, \label{EOM1} \\
{\cal P}_{\{ T^{-1}[C_1],..,T^{-1}[C_n] \}}^\HKP - {\cal P}_\CCC^\HK & = & 
\frac{\partial {\cal H}[J^\HK,P^\HKP] }{\partial J_\CCC^\HK }. \label{EOM2}
\eqa
The second equation needs a further explanation.
In general, there is no inverse for the map $T$
because not all loops in step $l$ survives in step $(l+1)$.
If there is no inverse for $C_i$, we simply define 
${\cal P}_{\{.., T^{-1}[C_i],.. \}}^\HKP = 0$.

To solve the equations of motion,
one needs two boundary conditions for each $\CCC$.
One condition is given by Eq. (\ref{eq:BC}).
The other condition should be implemented dynamically,
namely by minimizing the whole action over
all possible paths of $J_\CCC^\HK$ and $P_\CCC^\HK$
subject to Eq. (\ref{eq:BC}).
This is analogous to the problem of finding
the classical trajectory of a particle
where the initial position is fixed,
but the initial velocity is a variational 
parameter one uses to minimize the action.

\begin{figure}[h!]
\centering
        \includegraphics[height=2cm,width=10cm]{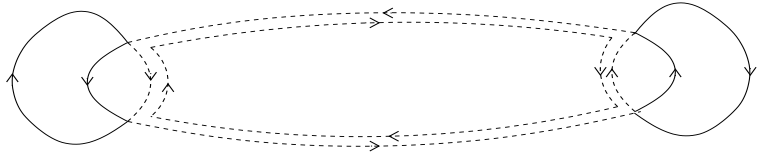}
\caption{
A non-local coupling between two separated circular small loops 
mediated by two large elongated loops.
The non-local interaction
creates dips in the final state of 
the small loops.
}
\label{fig:non_local_interaction}
\end{figure}

Is the theory local in $(D+1)$ dimensions ?
The theory is evidently local along the new dimension $l$.
The locality along the original $D$ dimensions is more tricky.
This is because size of loops can be arbitrarily large.
Even though one starts with small loops,
multi-loop interactions generate large loops.
Large loops can, in turn, mediate interactions between loops
which are far from each other.
Fig. \ref{fig:non_local_interaction} shows
an example where two large loops $C_1$ and $C_2$ mediate
interaction between two small loops $C_1^{'}$ and $C_1^{'}$ 
which are far from each other,
\bqa
P_{\{ L_1, L_2 \} }^\HKP
J_{C_1}^\HK J_{C_2}^\HK J_{C_1^{'}}^\HK J_{C_2^{'}}^\HK
\delta_{ C_{1,1}^{'} + C_{1,1}, T[L_1]}
\delta_{ C_{2,1}^{'} + C_{1,3}, T[L_2]}
\delta_{ C_{1,2} + C_{2,2}^{'} + C_{1,4} + C_{1,2}^{'}, C_2}.
\eqa
The non-local coupling between 
$J_{C_1^{'}}^\HK$
and 
$J_{C_2^{'}}^\HK$
is proportional to the amplitude
of the large loops, $J_{C_1}^\HK J_{C_2}^\HK$.
This looks bad for locality.
However, if the saddle point value of $J_C$ decreases exponentially
as the size of the loop increases,
the non-local coupling is exponentially small.
In this case, locality is still maintained.
%
Therefore, we can choose the initial couplings as
\bqa
J_{C_1}^{(0)} & \sim & \lambda^{-A_{C_1}}, \nn
J_{C_1,C_2}^{(0)} & \sim & \lambda^{-(A_{C_1}+A_{C_2} + d_{C_1,C_2})}, \nn
...
\label{eq:local}
\eqa
where $\lambda$ is the 't Hooft coupling much larger than $1$,
$A_C$ is the minimum area enclosed by the loop $C$,
and $d_{C,C^{'}}$ is the minimum distance between the two loops.
Now we prove that if the saddle point values of $J_\CCC^{(l)}$ 
are exponentially small for large loops,
those of $J_\CCC^{(l+1)}$ are also exponentially small for large loops.
Because the Hamiltonian depends on $P_\CCC$ only through
$r_\CCC$ in Eq. (\ref{eq:r}), 
Eq. (\ref{EOM1}) can be written as
\bqa
J_\CCC^\HKP - J_\DCCC^\HK & = & 
- \sum_\CCCT
\frac{\partial r_\CCCT^\HK }{\partial {\cal P}_\CCC^\HKP } 
\frac{\partial {\cal H} }{\partial r_\CCCT^\HK } \nn
& = &
-  \sum_\CCCT
h_{\DCCC;\CCCT}[J^\HK] 
 \frac{\partial {\cal H} }{\partial r_\CCCT^\HK },
\eqa
where we used 
\bqa
r_\CCCT^\HK =
J_\CCCT^\HK +
 \sum_{n=1}^\infty
\sum_\CCC
{\cal P}_\CCC^\HKP h_{\DCCC;\CCCT}[J^\HK]. 
\label{eq:r2}
\eqa
Note that $h_{\DCCC;\CCCT}[J^\HK]$ 
carries $Q^{ij}$ charges
for loops, $\DCCC$ and $\CCCT$.
Therefore, if any loop in $\CCC$ is large
there must be a large number of $J^\HK$'s
or at least one $J^\HK$ with a large loop 
in $h_{\DCCC;\CCCT}$ 
in order to match the charge.
Similarly, if there are two loops in $\CCC$
which are far from each other,
$h_{\DCCC;\CCCT}$ 
must include
either a large number of loop fields $J^\HK$ 
or at least one multi-loop fields such as $J_{C_1,C_2}^\HK$
where the separation between $C_1$ and $C_2$ is large
which connect the separated two loops.
This is because disconnected diagrams do not contribute to $h_{\DCCC;\CCCT}[J^\HK]$. 
This guarantees that if the condition in Eq. (\ref{eq:local})
is satisfied for a large $\lambda^\HK$ at time $l$,
a similar set of condition will be satisfied
with a large $\lambda^\HKP \sim \lambda^\HK$ at time $l+1$.
Therefore the theory will remain local
as far as the theory stays strongly coupled 
with large 't Hooft couplings along the 
renormalization group flow.
This is certainly true for small $l$ (UV region) 
if one starts with large 't Hooft couplings.

Whether the amplitudes for large loops remain exponentially
small for large $l$ (IR region) will be determined dynamically
through the equation of motion.
Here we can think of three possibilities.
The first possibility is that the energy dependent 't Hooft coupling $\lambda^\HK$
increases without a bound as $l$ increases.
In this case, the locality becomes better at IR.
On the other hand, fluctuations of loop fields become
suppressed and loops become very heavy in the IR limit.
This is the case for the pure $U(N)$ gauge theory
in the strong coupling limit $\lambda >> 1$.
In the low energy limit, there is no light degrees of freedom,
which describes the confinement phase.
The second possibility is the case where
$\lambda^\HK$ decreases to a small value 
in the IR limit.
If this happens, non-local couplings between loops
become significant below a certain energy scale ($l > l_c$).
In this case, the theory becomes non-local in the IR limit.
This occurs if the theory flows to a IR fixed point
which is not strongly coupled, that is, $\lim_{l \rightarrow \infty} \lambda^\HK  \leq 1$.
The last possibility, for which the holographic description is most useful,
is that the 't Hooft coupling stays 
at a large but finite value $\lim_{l \rightarrow \infty} \lambda^\HK  >> 1$ 
in the large $l$ limit.
This can be realized in a theory which flows
to a strongly interacting conformal field theory,
such as the ${\cal N}=4$ super Yang-Mills theory
at a large 't Hooft coupling.
Then the IR dynamics is described by
a weakly interacting local theory of closed loops.
At the saddle point, loop fields generically 
have nonzero expectation values.
Small fluctuations of loop fields around the 
classical configuration will be described by
a perturbative string theory.
However, LLFT is well defined even non-perturbatively,
and one can consider non-perturbative objects 
such as solitons.

\section{Discussions}

\subsection{Relation between LLFT and string field theory}

One may view LLFT as 
a string field theory put in lattice.
In order to obtain a 
string field theory in the smooth AdS space, 
it would be better to use 
a different regularization
for the gauge theory.
This is because RG steps can not be
continuous in the lattice regularized theory.
It would be of interest 
to devise a better regularization scheme
for strongly coupled gauge theory 
which allows one to construct a more cut-off independent 
holographic theory. 
However, the lattice construction will be more useful
to describe non-critical phases in real lattice models.

\subsection{Comparison with the $O(N)$ vector model}

The equation of motion in Eqs. (\ref{EOM1}) and (\ref{EOM2})
can be solved only if another boundary condition is provided
besides the UV boundary condition in Eq. (\ref{eq:BC}).
The extra boundary condition should be imposed dynamically in the IR limit.
In the case of the holographic $O(N)$ vector model\cite{SLEE10},
imposing the IR boundary condition
amounts to solving the original field theory.
Because the holographic theory for the $O(N)$ model has
non-singlet bulk fields which are not classical
even in the large $N$ limit,
one has to integrate over the non-singlet bulk fields 
in the IR limit
to dynamically impose the second boundary condition.
On the other hand, in the $U(N)$ gauge theory 
only $U(N)$ singlet fields are physical and 
all bulk degrees of freedom are classical
in the large $N$ limit. 
For this reason, the IR boundary condition can be imposed
by minimizing the whole action with respect to the
loop fields in the IR limit.
Whether this leads to a simple regularity condition
as in the standard AdS/CFT correspondence is yet to be understood.
Nonetheless, in large $N$ gauge theory, everything
boils down to solving classical problem,
which is the major difference from the $O(N)$ model.
Therefore, the duality between LLFT and the large $N$ gauge theory
is a strong-weak coupling duality,
contrary to the case for the $O(N)$ vector model.

\subsection{Outlook}

Here, we comment on some advantages and disadvantages 
of the present approach which is rather orthogonal to earlier ones.
First, the present prescription is general, and it can be applied to general quantum field theories.
Holographic duals constructed in this way are not expected to be very useful 
for quantum field theories for which the dual descriptions involve strong coupling or non-locality.
However, the strategy would be to establish a prescription applicable to general field theories, and then investigate how classical limit and locality emerge as dynamical features of certain theories.
Second, dual theories constructed in this prescription has full quantum actions in the bulk.
The correspondence is beyond the level of matching equations of motion in the bulk with beta functions of field theories.
Only in the large N limit, the classical equation of motion in the bulk can be directly compared to the beta function of the boundary theories.
This is an aspect in which the present approach clearly differs from the standard Wilsonian RG.
Third, the basic bulk degrees of freedom in this approach are fluctuating sources and vacuum expectation values of operators
in terms of which the standard AdS/CFT conjecture has been formulated.

The present approach also has some disadvantages.
First, the diffeomorphism invariance is not manifest.
Since the bulk space is emergent, it is expected that there should be a redundancy in parameterizing the space. 
Choosing a different gauge would amount to choosing a different prescription of real space RG, 
such as coarse graining degrees of freedom at different rates at different points in space.
In the present formalism,
the gauge redundancy is not manifest because specific `time' slices are chosen.
To make the diffeomorphism invariance more explicit, 
it may be useful 
to integrate in pure gauge degrees of freedom in the formalism based on continuous RG step.
Second, there is a large number of non-dynamical fields in the bulk.
In gauge theory, one has to keep infinitely many loop fields associated with 
multi-trace operators even for a finite $N$.
Although this issue is less acute in the large $N$ limit,
it would be interesting to explicitly remove non-dynamical fields from the beginning.
This is in contrast to the O(N) vector model where 
one needs to introduce only a finite number of fields
in the bulk\cite{SLEE10}.

\section{Acknowledgment}

I would like to thank
Alex Buchel,
Hong Liu,
Keith Ng,
Joao Penedones,
Eva Silverstein, 
Brian Swingle
and
Mithat Unsal
for helpful comments and discussions.
This research was supported in part by 
the Natural Sciences and Engineering Research Council of Canada
and the Early Research Award from the Ontario Ministry of Research and Innovation.
Research at the Perimeter Institute is supported 
in part by the Government of Canada 
through Industry Canada, 
and by the Province of Ontario through the
Ministry of Research and Information.

\end{document}